\documentclass[10pt, conference, letterpaper]{IEEEtran}

\pdfoutput=1

\usepackage{cite}
\usepackage[font=small]{caption}
\usepackage{subfig}
\usepackage{graphicx}
\usepackage{amssymb}
\usepackage{amsmath}
\usepackage{amsfonts}
\usepackage{mathrsfs}
\usepackage{mathtools}
\usepackage{bm}
\usepackage{bbm}
\usepackage{xspace}
\usepackage{balance}

\usepackage{url}

\usepackage{algorithm}
\usepackage{algpseudocode}
\usepackage{xpatch}
\xpatchcmd{\algorithmic}{\setcounter}{\algorithmicfont\setcounter}{}{}
\providecommand{\algorithmicfont}{}
\providecommand{\setalgorithmicfont}[1]{\renewcommand{\algorithmicfont}{#1}}
\algrenewcommand\algorithmicindent{1.5em}
\algrenewcommand\alglinenumber[1]{{\sffamily\footnotesize#1}}
\makeatletter
\xpatchcmd{\algorithmic}{\itemsep\z@}{\itemsep=1.5pt}{}{}
\makeatother
\algnewcommand\algorithmicinput{\textbf{Input:}}
\algnewcommand\Input{\item[\algorithmicinput]}
\algnewcommand\algorithmicoutput{\textbf{Output:}}
\algnewcommand\Output{\item[\algorithmicoutput]}

\usepackage{tikz}
\usetikzlibrary{
                arrows,
                plotmarks,
                patterns,
                fadings,
                shapes,
                shapes.multipart,
                positioning,
                graphs, 
                calc,
                matrix
}

\newcommand{\tabbedbox}[5]{
    \node[draw,
          ultra thin,
          rounded corners,                 
          rectangle split,
          rectangle split parts = 2,
          rectangle split horizontal,
          rectangle split part fill = {#2,none},
          every text node part/.style = {text=white, font=\bf\scriptsize\sffamily},
          rectangle split empty part width = #4,
          rectangle split empty part height = #5]
    (a) at (#1) {\rotatebox{90}{#3}\nodepart{two}{}};
    \fill[#2!30, rounded corners, fill opacity = 0.25] (a.south west) rectangle (a.north east);
}

\usetikzlibrary{external}
\tikzset{external/export=true}  

\usepackage{pgfplots}
\pgfplotsset{compat=newest,major grid style={dotted,line width=.5pt}}
\usepgfplotslibrary{groupplots}
\newlength\figureheight
\newlength\figurewidth

\usepackage{amsthm}

\begingroup
    \makeatletter
    \@for\theoremstyle:=definition,remark,plain\do{%
        \expandafter\g@addto@macro\csname th@\theoremstyle\endcsname{%
            \addtolength\thm@preskip\parskip
            }%
        }
\endgroup

\makeatletter
\newcommand*{\etc}{%
    \@ifnextchar{.}%
        {\textit{etc}}%
        {\textit{etc.}\@\xspace}%
}
\makeatother
\newcommand*{\eg}{\textit{e.g.},\@\xspace}
\newcommand*{\ie}{\textit{i.e.},\@\xspace}
\newcommand{\etal}{\textit{et~al}.~}
\newcommand{\eqnend}{\,.}
\newcommand{\eqncnt}{\,,}
\newcommand{\tth}{\textsuperscript{th}\@\xspace}
\newcommand{\expval}{\mathbb{E}}
\newcommand{\prob}[1]{\mathbbm{P}\left( #1 \right)}

\newcommand{\pmf}{{p.m.f.}\xspace}  
\newcommand{\iid}{{i.i.d.}\xspace}  

\newcommand{\fig}[1]{Fig.~\ref{#1}}

\newcommand{\eqn}[1]{Eq.~(\ref{#1})}
\newcommand{\eq}[1]{(\ref{#1})}

\mathchardef\mdash="2D

\begin{document}

\title{Characterizing Interest Aggregation in Content-Centric Networks}

\author{
\IEEEauthorblockN{
Ali Dabirmoghaddam$^*$\hspace{2em}Mostafa Dehghan$^\dagger$\hspace{2em}J. J. Garcia-Luna-Aceves$^{* \, \ddagger}$
}
\IEEEauthorblockA{
$^*$University of California Santa Cruz, 
$^\dagger$University of Massachusetts Amherst,
$^\ddagger$PARC\\
\texttt{\{alid, jj\}@soe.ucsc.edu, mdehghan@cs.umass.edu}
}
}

\maketitle

\begin{abstract}
  The Named Data Networking (NDN) and Content-Centric Networking (CCN)
  architectures advocate Interest aggregation as a means to reduce
  end-to-end latency and bandwidth consumption. To enable these
  benefits, Interest aggregation must be realized through Pending
  Interest Tables (PIT) that grow in size at the rate of incoming
  Interests to an extent that may eventually defeat their original
  purpose. A thorough analysis is provided of the Interest aggregation
  mechanism using mathematical arguments backed by extensive
  discrete-event simulation results. We present a simple yet accurate
  analytical framework for characterizing Interest aggregation in a
  CCN router, and use our model to develop an iterative algorithm to
  analyze the benefits of Interest aggregation in a network of
  interconnected routers. Our findings reveal that, under realistic
  assumptions, an insignificant fraction of Interests in the system
  benefit from aggregation, compromising the effectiveness of using
  PITs as an integral component of Content-Centric Networks.
\end{abstract}


\IEEEpeerreviewmaketitle

\section{Introduction} \label{sec:introduction}


The fact that the content and not its location is what matters to
end-users has given rise to many recent proposals classified under the
generic name of Information-Centric Networking
(ICN)~\cite{ahlgren:12}.
Many ICN blueprints can be seen as \emph{Interest-driven}
communication models, where users ask for content by name through
\emph{Interest} packets.  The most prominent examples of
Interest-driven ICN (NDN \cite{zhang:14} and CCNx \cite{ccnx}) are
based on three main components. Routers maintain \emph{Forwarding
  Information Bases} (FIB) listing routes to name prefixes, which
enable routing to \emph{names} rather than addresses. Given that all
copies of the same content are equivalent, routers cache content
opportunistically in their local \emph{Content Store} (CS).
In addition, each router maintains a \emph{Pending Interest Table}
(PIT) to suppress unnecessary Interests.

When a router receives an Interest that cannot be satisfied through
its local CS, it creates a PIT entry and forwards the Interest if
there is no entry for the same content name in its PIT.  If a PIT
entry already exists for the content name, the Interest is
\emph{aggregated} at the router and is not forwarded.
The idea of Interest aggregation is hardly new. It has been
implemented in Web caching architectures in the past, \eg
Squid~\cite{squid}---where it was referred to as \emph{collapsed
  forwarding}---and commercially used on production content delivery
networks since the early days of the Web.

The expectation of Interest aggregation in ICN architectures has been
that network and server loads can be drastically reduced by
suppressing similar Interests, and that end-to-end latencies can be
reduced by integrating caching with Interest aggregation.  These
benefits, however, come at a non-trivial cost. Creating and
maintaining the PIT is expensive, especially when performed at
Internet scale. The routers used in the Internet backbone handle
hundreds of thousands of packets every second. The proposed data
structure must be fast enough when operating at its peak capacity to
not act as a source of latency and overhead itself. Thus, considerable
work has focused on optimization and scalability of the PIT (\eg
see~\cite{dai:12,wang:12,varvello:13,yuan:14}).

We are not aware of any comprehensive analytical work characterizing
the expected benefits of Interest aggregation. Some experimental
efforts~\cite{virgilio:13,abu:14} have been made in understanding the
dynamics of the PIT size; however, important questions such as
\emph{what fraction of Interests are subject to aggregation under
  realistic conditions} and \emph{whether or not that justifies the
  use of PITs} have remained unanswered to this date.

To answer these questions, Section~\ref{sec:lru_model} presents a
simple yet powerful analytical toolbox for characterizing a CCN router
with a CS and a PIT. We compute the cache hit probability at the CS,
the Interest aggregation probability at the PIT, as well as the router
response time at an object-level
granularity. Section~\ref{sec:network_model} uses these constructions
for the analysis of a network of interconnected content
routers. Through extensive event-driven simulations,
Section~\ref{sec:simulations} demonstrates how accurately our proposed
framework can predict the steady-state behavior of such a complex
system. Furthermore, it shows how the model can be used to study the
performance of a large network under realistic conditions, such as
that of today's Internet, for which event-driven simulations are
prohibitive.

Numerical evaluations of our model for large-scale systems reveal that
only a small fraction of Interests may actually benefit from Interest
aggregation in realistic settings. These benefits are highly dependent
on the amount of caching budget available to the network. For example,
a 5\% cumulative aggregation percentage can be achieved using a
per-node caching capacity of equal to 0.05\% of the total number of
objects in the system, while increasing the budget to just 0.5\%
reduces the aggregation percentage to below 1\%. Even worse, our
findings show that most Interest aggregation takes place closer to
where the content is permanently stored---\ie near the producers deep
in the core of the network---where aggregating Interests hardly makes
sense anymore.

The insights from our modeling results lead to the necessary
conclusion that Interest aggregation should {\em not} be an integral
component of future ICN architectures and Content-Centric
Networks. On-path caching or edge caching provide all the benefits of
reducing the number of Interests that request similar content, without
the costs of maintaining PITs. In turn, if per-Interest forwarding
state is not needed for other reasons, this realization makes
Content-Centric Networking at Internet scale more feasible than the
current NDN design, given that forwarding data structures (\eg
CCN-DART~\cite{jj-icccn:15,jj-icnc:16} and CCN-GRAM \cite{jj-ifip:16})
smaller and more efficient than PIT could be used for routing data
back to the consumers.


\section{ CCN Router with Non-zero Download Delays}\label{sec:lru_model}

We develop a mathematical model to characterize a CCN router with a
Content Store (CS) to enable caching functionality and a Pending
Interest Table (PIT) that allows Interest aggregation. Unlike previous
work~\cite{dan:90,che:02,ioannidis:08,rosensweig:13}, we assume that
the content download delays are non-zero. Our derivations, hence, can
be regarded as an extension to a highly accurate approximation of LRU
cache introduced by Che \etal~\cite{che:02}.

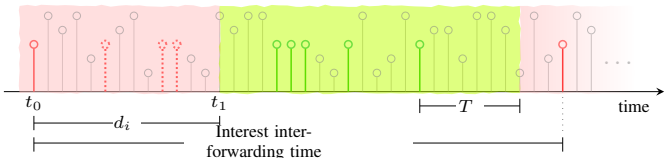
\begin{figure}[h]
  \centering
  \tikzset{external/export next=false}
  \pgfplotsset{
  standard/.style={
    axis x line=middle,
    axis y line=none,
    enlarge x limits=0.05,
    enlarge y limits=0.05,
    every axis plot post/.style={mark=o, mark size=1.5pt, mark options={fill=none, draw opacity=0.5,yshift=1.75pt, rotate=8}}
  }
}

\tikzsetnextfilename{cache_requests}
\begin{tikzpicture}[scale=.95]
\begin{axis}[%
  standard,
  x=2mm,
  y=2mm,
  ymin=-4,
  ymax=8,
  xmax=43,
  ticks=none,
  ]

  \node [font=\scriptsize] at (axis cs:1, -0.8) {$t_0$};
  \node [font=\scriptsize] at (axis cs:14,-0.8) {$t_1$};
  \node [font=\scriptsize] at (axis cs:43,-1) {time};
  \draw [ultra thin] (axis cs:1, -1.7) -- (axis cs:1, -2.7);
  \draw [ultra thin] (axis cs:14,-1.7) -- (axis cs:14,-2.7);
  \draw [ultra thin] (axis cs:1, -2.2) -- (axis cs:14,-2.2) node [midway,text width=1.1em,inner sep=0pt, text centered,font=\scriptsize,fill=white] {$d_i$};
  \draw [ultra thin] (axis cs:28,-0.5) -- (axis cs:28,-1.5);
  \draw [ultra thin] (axis cs:35,-0.5) -- (axis cs:35,-1.5);
  \draw [ultra thin] (axis cs:28,-1.0) -- (axis cs:35,-1.0) node [midway,text width=1.1em,inner sep=0pt,text centered,font=\scriptsize,fill=white] {$T$};
  \draw [ultra thin] (axis cs:1, -3.2) -- (axis cs:1, -4.2);
  \draw [ultra thin] (axis cs:38,-3.2) -- (axis cs:38,-4.2);
  \draw [ultra thin] (axis cs:1, -3.6) -- (axis cs:38,-3.6) node [midway,text width=9.1em,inner sep=0pt,text centered,font=\scriptsize,fill=white] {Interest inter-forwarding time};
  \draw [very thin, dotted] (axis cs:38, 0) -- (axis cs:38, -3);

  \addplot[ycomb, draw=red, semithick]
  table [
    ] {
    x y
    1 3
    38 3
  };

  \addplot[ycomb, draw=red, densely dotted, thick]
  table [] {
    6 3
    10 3
    11 3
  };

  \addplot[ycomb, draw=black!25!green, semithick]
  table [] {
    18 3
    19 3
    20 3
    23 3
    28 3
  };

  \addplot[ycomb, draw=gray, draw opacity=.5, very thin]
  table [
  ] {
    x y
    2 5
    3 4
    4 5
    5 2
    7 4
    8 5
    9 1
    12 2
    13 1
    14 5
    15 4
    16 5
    17 5
    21 2
    22 1
    24 5
    25 1
    26 2
    27 5
    29 4
    30 4
    31 2
    32 5
    33 5
    34 4
    35 1
    36 5
    37 2
    39 5
    40 4
  };

  \node [text=gray!50] at (axis cs: 42,2) {$\cdots$};

  \pgfmathsetseed{1357}
  \draw [draw=none, fill=pink, fill opacity=0.5, decorate, decoration={random steps,segment length=3pt,amplitude=0.5pt}] (axis cs:0, -0.1) rectangle (axis cs:14,6);
  \pgfmathsetseed{1357}
  \draw [draw=none, fill=lime, fill opacity=0.5, decorate, decoration={random steps,segment length=3pt,amplitude=0.5pt}] (axis cs:14,-0.1) rectangle (axis cs:35,6);
  \path [scope fading=east] (axis cs:35, -0.1) rectangle (axis cs:43, 6);
  \pgfmathsetseed{1357}
  \draw [draw=none, fill=pink, fill opacity=0.5, decorate, decoration={random steps,segment length=3pt,amplitude=0.5pt}] (axis cs:35,-0.1) rectangle (axis cs:43,6);

\end{axis}
\end{tikzpicture}

  \caption{Interests arriving at a content router over time 
  }
  \label{fig:cache_requests}
\end{figure}

Consider a content router with a CS of capacity $C$ implementing LRU
replacement policy receiving Interests indexed (without loss of
generality) in their decreasing order of popularity from 1 to $N$.
In the rest of this paper, the terms CS and cache refer to the
same concept and we shall use them interchangeably. Assume that
Interests conform to the Independent Reference Model (IRM), \ie for
every object, Interests inter-arrival times to the router are independent, 
identically distributed (\iid) random variables. \fig{fig:cache_requests} 
illustrates Interests (shown as combs) as arriving at a content router over 
time. We focus on the Interests for an object $i$, which are highlighted 
in color. In \fig{fig:cache_requests}, the red and green zones respectively 
specify time intervals when object $i$ is absent or present in the CS. Upon
receiving an Interest, if the CS contains a copy of object $i$, a
\emph{cache hit} occurs (see green combs); the Interest is immediately
satisfied and a copy of that object is sent back to the
requester. Otherwise, we say a \emph{cache miss} occurs (see red
combs).

Let $t_0$ be the instant when the first cache miss for object $i$
occurs at the CS. At that point, the content router creates a PIT
entry for that object and forwards the Interest to another
router/source (forwarded Interests shown as solid red combs). Let
$d_i$ be the random variable indicating the duration till a copy of
the requested object is downloaded and stored in the CS, and mark that
instant as $t_1 = t_0 + d_i$. We shall refer to $d_i$ as the
\emph{download delay} of object $i$. Any subsequent Interest for
object $i$ during the interval $(t_0, t_1)$ is aggregated at the
content router due to the existence of a PIT entry under object $i$'s
name (aggregated Interests shown as dotted red combs).
    
A set of events take place at time $t_1$, namely the content router:
(1) stores a copy of object $i$ in the local CS; (2) removes the PIT
entry for the corresponding Interest, and (3) forwards a copy of
object $i$ on all interfaces from which a request for it had been
received.  A copy of object $i$ remains in the CS so long as the
inter-arrival time of consecutive Interests for object $i$ is smaller
than $T$ denoting the \emph{characteristic time} of the cache
introduced by Che \etal~\cite{che:02} (see
\fig{fig:cache_requests}). In essence, $T$ is a random variable
signifying the duration it takes until $C$ distinct objects other than
$i$ are downloaded into the cache and object $i$ is dismissed. For
relatively large $C$ and $N$ and when the content download delays into
the cache are zero, Fricker \etal~\cite{fricker:12itc} proved that $T$
becomes deterministic. Later Dehghan \etal~\cite{dehghan:15} proved
this right even when download delays are non-zero.

The characteristic time $T$ depends on the cache capacity $C$, the
Interests arrival rate and the object popularity distribution and is
computed according to
\begin{equation}
\label{eq:sum_bernoullis}
  \expval \Bigg[ \sum_{i = 1}^N X_i \Bigg] = C \eqncnt
\end{equation}
where $X_i$ is the Bernoulli random variable indicating whether object
$i$ is present in the cache or not. \eqn{eq:sum_bernoullis} comes from
the fact that the cache has the capacity for $C$ objects. Note that
$\expval[X_i]$ equals the probability of object $i$ being present in
the cache, \ie the cache occupancy probability. With Poisson arrivals
and thanks to the PASTA property, the cache occupancy probability
equals the cache hit probability for any object $i$. Hence, we arrive
at
\begin{equation}
  \label{eq:expval_sum_bernoullis}
  \sum_{i = 1}^N h_i = C \eqnend
\end{equation}
where $h_i$ denotes the cache hit probability of object $i$. We shall
later use \eq{eq:expval_sum_bernoullis} as a constraint in computing
the cache hit probability of individual objects.

\subsection{Computing the Cache Hit Probability}

Under the assumption that request inter-arrival times are independent
exponential random variables, for a particular object $i$, the \pmf of
exactly $n_i = k$ cache hits is the probability of the event that the
first $k$ Interest inter-arrival times are smaller than $T$, while the
following Interest inter-arrival time is greater than $T$. This
probability can be formalized by the geometric distribution
%
$  
\prob{n_i = k} =$ $ \big(1 - e ^ {-\lambda_i T}\big) ^ k \, e ^ {-\lambda_i T} \eqncnt
$
%
and the expected number of cache hits is derived as
\begin{equation*}
\label{eq:exp_num_hits}
  \expval[n_i] = \sum_{k=0}^{\infty} k \, \prob{n_i = k} = e ^ {\lambda_i T} - 1 \eqnend
\end{equation*}
After forwarding a missed request for object $i$, if $\expval[d_i]$
denotes the expected time to download a copy of $i$ into the CS, the
expected number of missed requests during this interval would be
%
$  \expval[\bar{n}_i] = 1 + \lambda_i \expval[d_i]$,
%
of which one Interest is forwarded and the remaining
$\lambda_i \expval[d_i]$ are aggregated at the PIT. The expected total
number of Interests received for object $i$ during one such
inter-forwarding cycle is then
%
$  \expval[N_i] = \expval[n_i] + \expval[\bar{n}_i]$.
%
Consequently, the probability of a cache hit for object $i$ is derived
as
\begin{equation}
  \label{eq:hit_prob}
  h_i = \frac{\expval[n_i]}{\expval[N_i]} 
      = \frac{e ^ {\lambda_i T} - 1}{\lambda_i \expval[d_i] + e ^ {\lambda_i T}} \eqnend
\end{equation}
\eqn{eq:hit_prob} can be regarded as an extension to the LRU
approximation of Che \etal~\cite{che:02} where download delays can be
non-zero. In fact, setting download delays to zero simplifies
\eqn{eq:hit_prob} to their renown form of
$h_i = 1 - \exp(-\lambda_i T)$.

\subsection{Computing the Interest Aggregation Probability}

We now turn to computing the probability of Interest aggregation at
the PIT. From our previous discussion, the expected number of
aggregated requests during the download interval $\expval[d_i]$ are
$\expval[\bar{n}_i] - 1 = \lambda_i \expval[d_i]$. The probability of
an Interest for object $i$ being aggregated at the PIT is subsequently
derived as
\begin{equation}
\label{eq:agg_prob}
  a_i = \frac{\expval[\bar{n}_i] - 1}{\expval[N_i]} 
      = \frac{\lambda_i \expval[d_i]}{\lambda_i \expval[d_i] + e ^ {\lambda_i T} } \eqnend
\end{equation}
Put differently, \eqn{eq:agg_prob} states what
fraction of Interests for object $i$ arriving at the content router are
aggregated in the long run.

\subsection{Computing the Router Response Time}

Another important measure in the analysis of interconnected routers
when the download delays are non-zero is the router response time. Due
to the potential existence of a PIT entry in the content router when an
Interest for object $i$ is received, the time it takes the router to
satisfy that Interest can be any value from interval $(0, d_i]$.

We define the \emph{pending time} of an Interest in the PIT as the
time difference between the arrival of the Interest to the router and
the subsequent moment when the Interest is served. With Poisson
arrivals, Interest arrival times are uniformly distributed over
$(0, d_i]$; hence, the sum $W_i$ of pending time of Interests during a
download interval $d_i$ can be formulated as
\[
W_i = d_i + \lambda_i \int_0^{d_i} (d_i - t) \, \mathrm{d}t 
        = d_i (1 + 0.5 \lambda_i d_i) \eqnend
\]
%
We define the response time $r_i$ of the content router for a particular
object $i$ as the expected pending time of Interests for that object
which is readily derived as
\begin{equation}
  \label{eq:resp_time}
  r_i = \frac{\expval[W_i]}{\expval[N_i]} 
      = \frac{\expval[d_i (1 + 0.5 \lambda_i d_i)]}{\lambda_i \expval[d_i] + e ^ {\lambda_i T}} \eqnend
\end{equation}

The router response time depends on the distribution of download
delays, though knowledge of only the first two moments is sufficient.


\section{An Algorithm for the Analysis of Hierarchical CCN Networks} \label{sec:network_model}

\begin{figure}[b]
  \centering
  \tikzsetnextfilename{network_tree}
\begin{tikzpicture}[scale = 0.65]
    \tikzset{
      every node/.style = {scale = 0.8},
      node distance     = 1.9em,
      cache/.style      = {minimum size=10pt,inner sep=0pt,circle,draw,scale=0.75},
      dots/.style       = {draw=none,scale=0.75},
      annot/.style      = {midway,sloped,text width=1.5em,inner sep=0pt,text centered,font=\scriptsize,fill=white},
      cons/.style       = {minimum size=2pt,inner sep=0pt,cross out,draw,scale=0.9},
      arrow/.style      = {->,>=stealth'},
      link/.style       = {scale=0.75},
    }

 
    \node [dots]  (d01)   at (2   ,0.2) {$\ldots$};
    \node [dots]  (d03)   at (3.75,0.2) {$\ldots$};
    \node [dots]  (d03)   at (6.5 ,0.2) {$\ldots$};
    \node [dots]  (d11)   at (1.5,1)  {$\cdots$};
    \node [cache] (c11)   at (3   ,1) {};
    \node [dots]  (d12)   at (3.75,1) {$\cdots$};
    \node [cache] (c12)   at (4.5 ,1) {};
    \node [cache] (c13)   at (5   ,1) {};
    \node [cons, below of=c11](inv1) {};
    \node [cons, below of=c12](inv2) {};
    \node [cons, below of=c13](inv3) {};
    \node [dots]  (d21)   at (2   ,2) {$\cdots$};
    \node [cache] (c21)   at (3.75,2) {};
    \node [dots]  (d22)   at (5.5 ,2) {$\cdots$};
    \node [cache] (c31)   at (0   ,3.5) {};
    \node [dots, below of=c31] (vd3)    {$\vdots$};
    \node [cache] (c32)   at (1   ,3.5) {};
    \node [dots]  (d31)   at (2   ,3.5) {$\cdots$};
    \node [cache] (c33)   at (4   ,3.5) {};
    \node [dots]  (d32)   at (5 ,3.5)   {$\cdots$};
    \node [cache] (c41)   at (2   ,4.5) {};
    \node [cache] (c42)   at (3   ,4.5) {};
    \node [dots]  (d4)    at (4   ,4.5) {$\cdots$};
    \node [cache] (c43)   at (5   ,4.5) {};
    \node [cache] (c44)   at (6   ,4.5) {};
    \node [dots, below of=c44, yshift=0.77em]  (vd4) {$\vdots$};
    \node [cache,fill=black]
                  (c51)   at (4   ,5.5) {};

    \draw [link]  (c51)  -- (c41);
    \draw [link]  (c51)  -- (c42);
    \draw [link]  (c51)  -- (c43);
    \draw [link]  (c51)  -- (c44);
    \draw [link]  (c41)  -- (c31);
    \draw [link]  (c41)  -- (c32);
    \draw [link]  (c41)  -- (c33);
    \draw [link]  (c41)  -- (c21) node [annot] {$\cdots$};;
    \draw [link]  (c21)  -- (c11);
    \draw [link]  (c21)  -- (c12);
    \draw [link]  (c21)  -- (c13);
    \draw [arrow] (inv1) -- (c11);
    \draw [arrow] (inv2) -- (c12);
    \draw [arrow] (inv3) -- (c13);

    \tabbedbox{2,0.2} {black} {\rotatebox{-90}{\makebox[6em]{consumers}}}            {18em}{0.18em}
    \tabbedbox{2,1}   {gray}  {\rotatebox{-90}{\makebox[6em]{$\ell_1$ routers}}}      {18em}{1.25em}
    \tabbedbox{2,2}   {gray}  {\rotatebox{-90}{\makebox[6em]{$\ell_2$ routers}}}      {18em}{1.25em}
    \tabbedbox{2,3.5} {gray}  {\rotatebox{-90}{\makebox[6em]{$\ell_{L - 1}$ routers}}}{18em}{1.25em}
    \tabbedbox{2,4.5} {gray}  {\rotatebox{-90}{\makebox[6em]{$\ell_L$ routers}}}      {18em}{1.25em}
    \tabbedbox{2,5.5} {black} {\rotatebox{-90}{\makebox[6em]{producer}}}            {18em}{1.25em}

\end{tikzpicture}

  \caption{A partial view of a hierarchy of interconnected routers 
  }
  \label{fig:network_tree}
\end{figure}
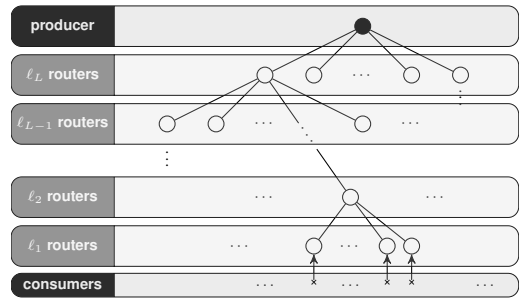

We investigate how an interconnected network of CCN routers can be
analyzed using the results from the previous section. Consider a
hierarchy of routers as depicted in \fig{fig:network_tree}. Consumers
are located at the bottom level where their requests for objects of
interest are directed to the first level CCN routers (\ie $\ell_1$
routers). An $\ell_1$ router searches its local CS for a copy of the
requested object and if failed, it forwards the request to the next
level router (\ie the parent $\ell_2$ router). This process is
repeated on every cache miss and in the worst case, the requested
object is downloaded directly from the producer at the top of the
hierarchy storing permanent copies of all the objects in the
system. On the reverse path back to the original requester, a copy of
the object is stored in the CS of every CCN router it passes through.

For simplicity, we consider only a single producer in the network.
The producer in our model can alternatively be conceived as a
collection of several producers at the core of the network collapsed
into one single entity. This single-source spanning-tree
simplification of the network topology is standard in many studies of
content delivery~\cite{ni:05} and publish-subscribe
networks~\cite{chand:04}.

There are two important challenges in the analysis of the foregoing
structure. First, the Interest stream into a higher-level router (\ie
all except $\ell_1$ routers) is no longer a simple Poisson process,
but an aggregate of miss streams from a number of lower-level
routers. It is known, however, that the superposition of multiple
streams tends toward Poisson as the load
increases~\cite{cao:03,cinlar:68}. We shall use this insight when
extending the results of the previous section to the analysis of CCN
networks by primarily focusing on trees of higher arity.

Secondly, chaining routers may cause circular dependencies in the
computation of some router performance metrics. For instance, the
cache hit probability in an $\ell_1$ router depends on the download
delay of the objects as mandated by \eqn{eq:hit_prob}. That delay is
determined by the response time of the parent $\ell_2$ router which in
turn is a function of its input rate. The input into an $\ell_2$
router itself partially relies on the miss stream of its descendant
$\ell_1$ router, and that is how a dependency cycle is formed. To
overcome this hurdle, we present an iterative approach as outlined in
Algorithm~\ref{alg:cache_hierarchy}.

Procedure {\sc Analyze-CCN-Tree} is proposed to compute the important
router performance metrics we discussed in Section~\ref{sec:lru_model}
for a hierarchical network structure such as the one portrayed in
\fig{fig:network_tree}. The procedure analyzes a complete $k$-ary tree
of height $L+2$ in which consumers are at level $0$; $L$ layers of CCN
routers are employed in the middle, and the producer is located at
level $L+1$ as the root of the tree. The available caching budget is
provided by vector $\bm{C}$ in which element $C_\ell$ indicates the
allocated CS capacity for each of the routers at level $\ell$. The
initial rate at which Interests are produced by the consumers and fed
into each $\ell_1$ router is $\lambda$. The object popularity profile
follows a Zipfian distribution as determined by the probability vector
$\bm{q}$. Without loss of generality, in this paper we always rank
objects in their decreasing order of popularity. As such, the
normalized popularity of the $n$\tth ranked object is determined by
the power-law
%
 $ q(n) = n^{-\alpha} / \sum_{u = 1}^N u^{-\alpha} \eqncnt$
%
where exponent $\alpha > 0$ is the parameter to the Zipf
distribution. Finally, each link induces a round-trip delay of
$\delta$ for transporting an individual content object. These
parameters are inputs to Algorithm~\ref{alg:cache_hierarchy}.

\setalgorithmicfont{\footnotesize}
\begin{algorithm}[t]
  \caption{Method to characterize a hierarchical CCN represented by a
    complete tree as depicted in \fig{fig:network_tree}}
\label{alg:cache_hierarchy}
\begin{algorithmic}[1]
\Input{
  $k$: arity of the tree;
  $L$: number of tree levels;
  $\lambda$: consumer input rate to each first level router;
  $\delta$: round-trip delay across each link;
  $\bm{C}$: vector of caching budget per node per layer;
  $\bm{q}$: probability vector reflecting the object popularity profile.
}
\Output{
  $T$: characteristic time of caches at each level;
  $\bm{h}$: vector of cache hit probabilities at each level;
  $\bm{a}$: vector of aggregation probabilities at each level;
  $\bm{r}$: vector of router response times at each level;
  $\bm{m}$: vector of incoming Interest rates to each level.
}
\Procedure{Analyze-CCN-Tree}{$k, L, \lambda, \delta, \bm{C}, \bm{q}$}
\State $i \gets 0$
\For{$\ell$ from 1 to $L$}                                                  \label{algln:beg_init_resp_time}
    \State $\bm{r}_{\ell + 1}^{(i)} \gets \delta \times (L - \ell)$
\EndFor                                                                     \label{algln:end_init_resp_time}
\While{not converged}                                                       \label{algln:while_beg}
    \State $i \gets i + 1$                                                  \label{algln:update_iter}
    \For{$\ell$ from 1 to $L$}                                              \label{algln:beg_update_delay}
        \State $\bm{d}_\ell^{(i)} \gets$ $\delta + \bm{r}_{\ell + 1}^{(i - 1)}$
    \EndFor                                                                 \label{algln:end_update_delay}
    \State $\bm{m}_1^{(i)} \gets \lambda \times \bm{q}$                      \label{algln:miss_rate_first}
    \For{$\ell$ from 1 to $L$}                                              \label{algln:for_beg_all_measures}
        \State \( \mathrlap{T_\ell^{(i)}}         \phantom{mml} \) $\gets$ \Call{Char-Time}{$\bm{m}_\ell^{(i)}$, $\bm{d}_\ell^{(i)}$, $C_\ell$}               \label{algln:char_time}
        \State \( \mathrlap{\bm{h}_\ell^{(i)}}    \phantom{mml} \) $\gets$ \Call{Hit-Prob}{$\bm{m}_\ell^{(i)}$, $\bm{d}_\ell^{(i)}$, $T_\ell^{(i)}$}           \label{algln:hit_prob}
        \State \( \mathrlap{\bm{a}_\ell^{(i)}}    \phantom{mml} \) $\gets$ \Call{Agg-Prob}{$\bm{m}_\ell^{(i)}$, $\bm{d}_\ell^{(i)}$, $T_\ell^{(i)}$}           \label{algln:agg_prob}
        \State \( \mathrlap{\bm{r}_\ell^{(i)}}    \phantom{mml} \) $\gets$ \Call{Resp-Time}{$\bm{m}_\ell^{(i)}$, $\bm{d}_\ell^{(i)}$, $T_\ell^{(i)}$}          \label{algln:resp_time}
        \State \( \mathrlap{\bm{m}_{\ell+1}^{(i)}}\phantom{mml} \) $\gets$ \Call{Miss-Rate}{$k$, $\bm{m}_\ell^{(i)}$, $\bm{h}_\ell^{(i)}$, $\bm{a}_\ell^{(i)}$} \label{algln:miss_rate}
    \EndFor                                                                 \label{algln:for_end_all_measures}
\EndWhile                                                                   \label{algln:while_end}
\EndProcedure
\end{algorithmic}
\end{algorithm}

As pointed out, Algorithm~\ref{alg:cache_hierarchy} works in
iterations to tackle circular dependencies. The superscript $(i)$ used
throughout the algorithm denotes the latest count of iterations. At
the $0$\tth iteration, \ie the initial phase, since all caches are
empty and all requests are fulfilled directly by the producer, the
router response times (denoted by $\bm{r}$) are simply set based on
the hop-distance of routers from the root
(lines~\ref{algln:beg_init_resp_time}--\ref{algln:end_init_resp_time}). The
notation $\bm{r}_{\ell+1}^{(i)}$ describes the response time of a
$(\ell+1)$\tth level router computed at the $i$\tth iteration. Note
that variables denoted in bold face are in fact vectors with values
corresponding to individual objects in the system as ordered in
popularity profile $\bm{q}$.
Next, at any subsequent iteration:
\begin{enumerate}
\item Download delays for all levels are updated
  (lines~\ref{algln:beg_update_delay}--\ref{algln:end_update_delay})
  according to the heuristic that the delay for downloading files into
  the CS of an arbitrary router is equal to the response time of its
  parent router plus the round-trip delay of the link connecting them
  together. Assuming all objects are unit-sized, we can deduce that
  the average link delays are the same. In a hierarchical structure,
  thus, we can compute the download delays into a particular router by
  knowing the average link delays and the response time of the next
  level (\ie parent) router.
\item All performance measures discussed in
  Section~\ref{sec:lru_model} are computed/updated across all tree
  levels
  (lines~\ref{algln:for_beg_all_measures}--\ref{algln:for_end_all_measures}).
\end{enumerate}

Starting from the bottom working towards the top, at each tree level
the measures are computed in the following order:

\paragraph{Procedure {\sc Char-Time}} is called at
line~\ref{algln:char_time} to compute the cache characteristic times
by solving the following fixed-point equation for variable $T$:
\begin{equation}
  \label{eq:chartime_fixedpoint}
  \sum_{j=1}^{N} \frac{e ^ {m[j] T} - 1}{m[j] d[j] + e ^ {m[j] T}} = C \eqncnt
\end{equation}
where $m[j]$ and $d[j]$ are the $j$\tth elements in vectors $\bm{m}$
and $\bm{d}$, respectively denoting the input rate and download delay
for object $j$ at the corresponding router. Note that
\eqn{eq:chartime_fixedpoint} is indeed the expanded form of
\eq{eq:expval_sum_bernoullis} using \eq{eq:hit_prob}.

\paragraph{Procedures {\sc Hit-Prob}, {\sc Agg-Prob} and {\sc
    Resp-Time}} are called at
lines~\ref{algln:hit_prob}--\ref{algln:resp_time} to use the above
computed characteristic time for computing the cache hit probability,
PIT aggregation probability and response time of routers for
individual objects according to Eqs. \eq{eq:hit_prob},
\eq{eq:agg_prob} and \eq{eq:resp_time}, respectively.

\paragraph{Procedure {\sc Miss-Rate}} is called at
line~\ref{algln:miss_rate} to compute the aggregate miss rate
\emph{into} the next level (\ie parent) router using the above
computed hit- and aggregation probabilities according to the following
relation:
\begin{equation}
  \label{eq:miss_rate}
  \bm{m}_{\ell+1} = k \cdot \bm{m}_{\ell} 
                     \odot (\bm{1} - \bm{h}_{\ell}) 
                     \odot (\bm{1} - \bm{a}_{\ell}) 
                     \eqncnt
\end{equation}
where $\odot$ signifies component-wise multiplication of corresponding
vectors. In essence, \eqn{eq:miss_rate} suggests that the input stream
of a router at level-$(\ell+1)$ is the superposition of $k$ miss
streams from its descendant level-$\ell$ routers. The only exception
are $\ell_1$ routers whose input is directly provided by consumers
according to line~\ref{algln:miss_rate_first}.

\begin{figure}[h]
  \centering
  \includegraphics{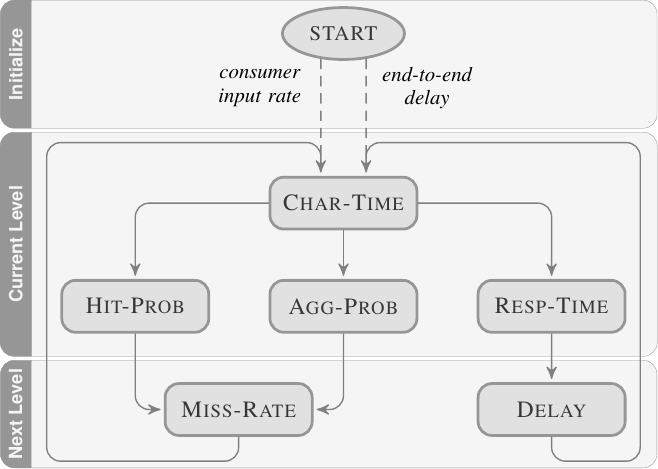}
  \caption{Dependency among procedure calls in
    Algorithm~\ref{alg:cache_hierarchy}.}
  \label{fig:flowchart}
\end{figure}

To better understand the dependency between these procedure calls, the
diagram in \fig{fig:flowchart} provides a pictorial view of their
relationship. At the beginning, consumers' input rate and the
end-to-end delay (between the consumers and the producer) are used to
compute the cache characteristic time for all $\ell_1$ routers. Cache
hit- and PIT aggregation probabilities as well as router response
times are then computed for $\ell_1$ routers. Next these results are
used to calculate the input rate and download delays for $\ell_2$
routers. Then level 2 becomes the current level and a similar
procedure is repeated for all remaining levels from the bottom to the
top of the tree.

The computations in the middle and bottom boxes in \fig{fig:flowchart}
may be repeated in consecutive iterations as needed; the results from
one iteration will be used in computing the next as the computed
measures gradually converge to their steady-state values. In our
numerical simulations---to be discussed next---we noticed that the
first few iterations usually suffice to get an accuracy of better than
$0.1\%$ while no more than 10 iterations were needed in all cases
studied (irrespective of the input size).

Implementing Algorithm~\ref{alg:cache_hierarchy} is straightforward in
many off-the-shelf numerical computing environments. In our
simulations, for solving \eqn{eq:chartime_fixedpoint}, we leveraged
\texttt{fsolve} function from MATLAB's Optimization Toolbox which uses
trust-region methods~\cite{conn:00} for solving systems of nonlinear
equations. It is known~\cite{gratton:08} that trust-region methods
take $\mathcal{O}(\epsilon^{-2})$ iterations to drive the norm of the
gradient of the objective function below desired threshold
$\epsilon$. The time-complexity of Algorithm~\ref{alg:cache_hierarchy}
is hence $\mathcal{O}(N L \, \epsilon^{-2})$.


\section{Performance Evaluation} \label{sec:simulations}

\begin{figure*}[t]
  \centering
  \setlength\figureheight{.152\textwidth} 
  \setlength\figurewidth{.2\textwidth}
  \input{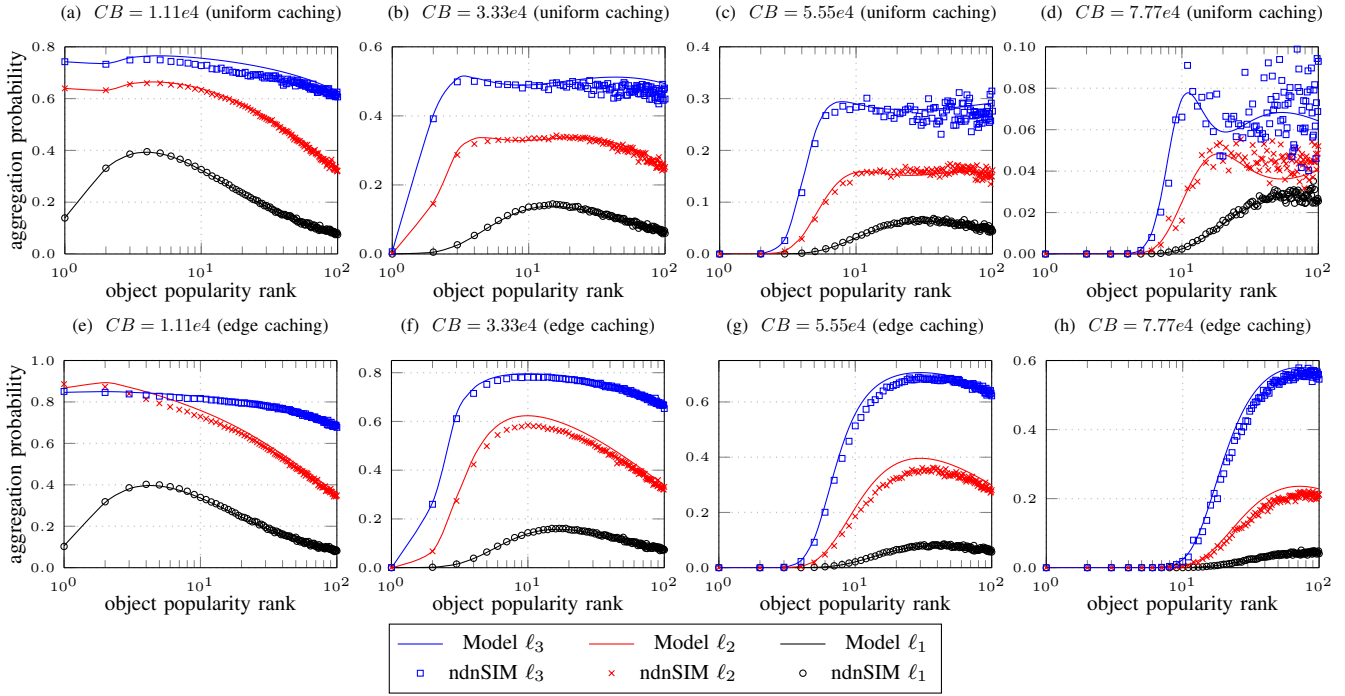}
  \caption{A comparison of model versus event-driven simulations. Input 
    rate into each edge router is 100 Interests/sec. Model predicts 
    aggregation probability for individual objects fairly accurately 
    across all levels of the tree topology. 
  }
  \label{fig:aggprob_detailed}
\end{figure*}

\begin{figure*}[ht]
  \centering
  \setlength\figureheight{.152\textwidth} 
  \setlength\figurewidth{.21\textwidth}
  \input{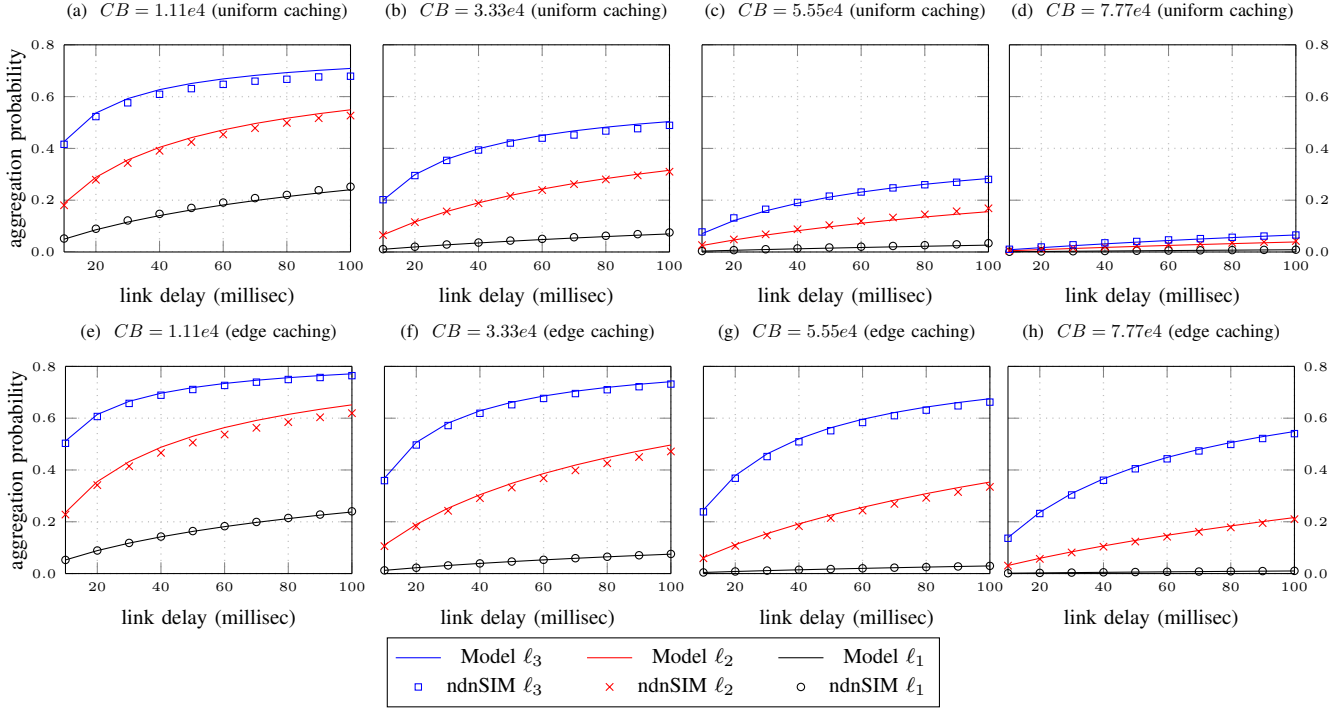}
  \caption{Interest aggregation probability at various router levels as
    a function of link delay for increasing cache sizes (left to
    right) and different cache allocation strategies (top vs bottom
    rows). Input rate into each edge router is 100 Interests/sec.
  }
  \label{fig:aggprob_delay}
\end{figure*}

We present simulation results to show how the proposed method can be
used to accurately predict the complex behavior of a network of content
routers. First, a detailed comparison of the numerical results of the
presented model versus the results from extensive event-driven
simulations in ndnSIM~\cite{ndnSIM} is presented. Next, the results
from our model are used to analyze more complex scenarios, such as
networks with larger content base, which are far more cumbersome and
time-consuming to study using conventional event-driven simulations.

We focus on two major strategies of cache allocation, namely
\emph{uniform caching} and \emph{edge caching}. In the former, a fixed
caching budget is evenly distributed across all content routers,
whereas with the latter, the budget is entirely allocated to the
routers at the edge of the network, \ie the ones directly serving the
consumers. With edge caching, the upper level routers simply act as
routers with no caching capability (that is, their CS size is set to
zero). Yet they still perform Interest aggregation upon receiving
Interests for which they have pending entries in their PITs.

\subsection{Comparison of Model with Event-driven Simulations}

We consider a tree of degree $k = 10$ and height $H = 5$ as the
underlying topology, where $L = 3$ levels of content routers are used in
the middle. The reason for using such a configuration is to keep the
overall aggregate traffic pattern in the middle layers as close to
being Poisson as possible, as discussed in
Section~\ref{sec:network_model}. Although the model was able to
capture the overall trends in our experiments with trees of lower
arity, we noticed that more accurate results are generally obtained
when nodes have higher fan-in (\eg 10 or more). This assumption,
however, is not unrealistic as some studies~\cite{caida} of the actual
Internet router-level topology have reported an average degree of more
than 22 per router.

For the first set of experiments, we begin with a fairly small content
catalog comprising only 100 objects. The reason for this choice is as 
follows. When
performing event-driven simulations with a large content base, the
system takes much longer to come to a steady-state; while growing
worse with an increasing caching budget. In such case, a large number
of requests must be used just to ``warm-up'' the system---hence, not
to be used for collecting statistics---from the initial state where
all caches are empty. Besides, because of the Zipfian nature of object
popularity, a larger number of requests must be generated in total to
ensure that objects at the long tail of the distribution also get a
reasonable chance to appear in the generated request stream. Even with
a content catalog of size 100 objects with a Zipf parameter of 1, we
had to generate roughly 4 million requests---while disregarding the
first half---to make sure all caches in all levels have their capacity
almost fully utilized before collecting statistics.

For the foregoing set-up, in \fig{fig:aggprob_detailed} we compare the
aggregation probability for individual objects as predicted by model
versus the results from extensive event-driven simulations. Curves in
each plot represent the PIT aggregation probability as attained by
\emph{each} of the content routers at the corresponding level. Thanks to
the symmetry of the topology, all routers at the same level share
similar statistics. Graphs in the top row contrast uniform caching
against edge caching employed for graphs at the bottom. In each row,
the total caching budget ($CB$) increases from left to right.
The model accurately predicts aggregation across
various caching levels even at a fine object-scale resolution. Edge
caching results in higher aggregation probability at higher
levels. This behavior is expected because with edge caching naturally
no cache hit may occur at higher levels in the tree. Therefore, many
requests that would have hit those caches if a non-zero cache size
were used will now end up being aggregated at PITs.

\begin{figure*}[ht]
  \centering
  \setlength\figureheight{.19\textwidth} 
  \setlength\figurewidth{.24\textwidth}
  \input{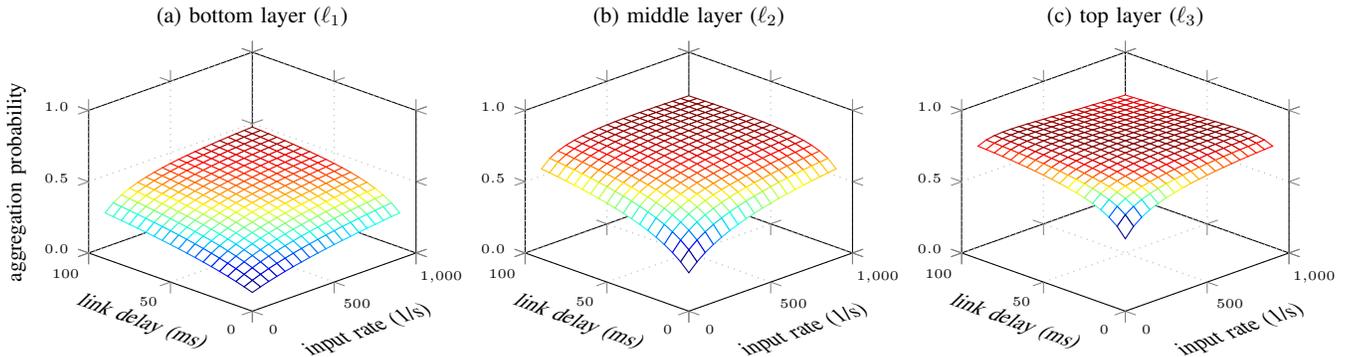}
  \caption{The combined impact of link delay and input rate on
    aggregation probability. Increasing one factor has exactly the
    same effect on Interest aggregation as increasing the other.}
  \label{fig:aggprob_delay_rate}
\end{figure*}

To obtain a more insightful view of Interest aggregation, graphs in
\fig{fig:aggprob_delay} show the odds of a generic Interest
(irrespective of the object popularity rank) getting aggregated at
each level of the tree when the link delay gradually increases. It is
clear that at a fixed Interest rate, an increased link delay generally
improves the aggregation probability. However, larger cache sizes tend
to offset some of these improvements, especially with uniform caching
strategy.

Interest aggregation occurs at a higher probability at upper levels of
the tree. This can be attributed to the higher input rate into those
levels considering the fact that the aggregate miss stream from many
lower level routers constitutes the input of their parent
router. Results from \fig{fig:aggprob_delay_rate} suggest that
significant benefits are likely to accrue from Interest aggregation;
however, this promising gain should be taken with a grain of salt due
to the reasons discussed in the following.

First, the small object catalog consisting of only 100 objects
naturally gives rise to a higher frequency of similar Interests
arriving at the router, thereby an increased aggregation
probability. Despite the event-driven simulation which turns out to be
extremely tedious especially for a large number of objects, numerical
simulations using the proposed model are practicable even on commodity
hardware. Our numerical results in the next subsection confirm that
the actual benefits of Interest aggregation are indeed much less in
reality.

Secondly, the notion of aggregation probability itself may give a
magnified image of the real benefits. In fact, aggregation probability
at a certain level in the hierarchy indicates what fraction of
Interests making it up to that level end up getting aggregated. Since
the request stream observed by the higher level routers is a
``filtered'' version of the input stream to their descendants, it is
clear that fewer Interests are received in total towards the top of
the hierarchy. For this, we define a new measure called
\emph{aggregation percentage} that determines the percentage ratio of
the count of aggregated Interests at a certain level (or at a
particular router) over the total count of produced Interests in the
whole system. Since every generated Interest can be aggregated at most
once on its path towards the producer, aggregation percentage provides
a more reasonable and unbiased measure, and we shall use that in our
later assessments of aggregation benefits.

Given the foregoing remarks, we emphasize that the results
demonstrated in Figs.~\ref{fig:aggprob_detailed} and
\ref{fig:aggprob_delay} are particularly meant to verify the accuracy
of the proposed analytical framework, and to provide a side-by-side
comparison of how varying different parameters affects the relative
odds of Interest aggregation. The true benefits of Interest
aggregation are discussed in the following subsection, where more
realistic input parameters are used.

\subsection{Numerical Evaluations}


\begin{table}[b]
  \centering
  \caption{Table of default parameter values}
  \begin{tabular}[h]{lll}
    \hline
    Parameter                        & Symbol    & Value           \\
    \hline
    Tree height                      & $H$       & 5               \\
    Number of cache layers           & $L$       & 3               \\
    Node degree                      & $k$       & 10              \\
    Total number of objects          & $N$       & 140 million     \\
    Cache capacity per cache node    & $C$       & 100,000 objects \\
    Zipf exponent                    & $\alpha$  & 0.8             \\
    Input rate into each edge cache  & $\lambda$ & 100,000/sec     \\
    Link delay each way              & $d$       & 15 milliseconds \\
    \hline
  \end{tabular}
  \label{tab:parameters}
\end{table}

\fig{fig:aggprob_delay_rate} sheds light on the combined impact of
download delay and input rate on the Interest aggregation
probability. The symmetry of plots in \fig{fig:aggprob_delay_rate}
suggests that it is in fact the combination of the link delay and
input rate which regulates the overall trend of aggregation
probability. In fact, doubling the input rate for a fixed link delay
has the same effect on the aggregation probability as keeping the
input rate fixed and doubling the link delay. Therefore, we define
\emph{system load} as the product of these two quantities to build our
next set of experiments on it. As a combined metric, system load does
not identify a specific delay or input rate, rather defines an
infinite range for these parameters. For example, a system load of 10
may imply an input rate of 100 Interests/sec with link delay of 0.1
seconds, or equivalently, an input rate of 500 Interests/sec with link
delay of 0.02 seconds.


In the experiments to be discussed next, we consider the same tree
topology as in the previous subsection, with the general
configurations listed in Table~\ref{tab:parameters} (unless otherwise
stated). The total number of objects considered, \ie 140 million, is
an estimation of the total number of videos on YouTube in
2008~\cite{russakovskii:08} and the Zipf parameter of 0.8 is taken
from empirical studies~\cite{mahanti:00,fricker:12infocom} of real
content networks. The input rate of 100,000 Interests/sec and link
delay of 15 milliseconds are also chosen such that the average
generated traffic in the network is comparable with the load
experienced by the Internet's backbone
routers~\cite{cowie:99,fraleigh:02}.

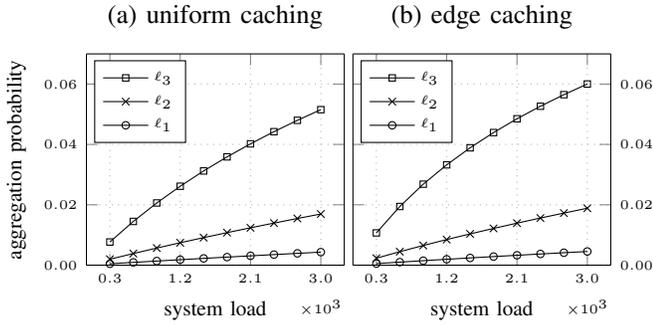
\begin{figure}[h]
  \centering
  \setlength\figureheight{.155\textwidth} 
  \setlength\figurewidth{.189\textwidth}
%
%
%
%
\tikzsetnextfilename{aggprob_sysload_uniform}
\begin{tikzpicture}

\begin{axis}[%
height=\figureheight,
width=\figurewidth,
at={(0.74in,0.469in)},
scale only axis,
separate axis lines,
every x tick label/.append style={font=\tiny, /pgf/number format/.cd, fixed, fixed zerofill, precision=1, /tikz/.cd},
every axis label/.style={font=\footnotesize},
xmajorgrids,
ymajorgrids,
xtick={300,1200,...,3000},
scaled x ticks=base 10:-3,
tick scale binop=\times,
every x tick scale label/.style={at={(axis cs:3000,-0.014)}},
xmin=0,
xmax=3300,
xlabel={system load},
every outer y axis line/.append style={black},
every y tick label/.append style={font=\tiny, /pgf/number format/.cd, fixed, fixed zerofill, precision=2, /tikz/.cd},
ytick={0,0.02,...,0.1},
ymin=0,
ymax=0.07,
scaled y ticks = false,
ylabel={aggregation probability},
axis background/.style={fill=white},
title={(a) uniform caching},
reverse legend,
legend style={font=\tiny},
legend pos=north west,
]
\addplot[draw=black,mark=o,mark size=1.4] plot table[row sep=crcr] {%
300	0.000468666119576182\\
600	0.000926376654977257\\
900	0.00137469003069925\\
1200	0.00181476571716393\\
1500	0.00224749633161581\\
1800	0.00267358857934827\\
2100	0.00309361518695588\\
2400	0.00350804952834169\\
2700	0.00391728947538705\\
3000	0.00432167429265729\\
};
\addplot[draw=black,mark=x] plot table[row sep=crcr] {%
300	0.00196845663967797\\
600	0.00385611151247761\\
900	0.0056738101149303\\
1200	0.00742971180750655\\
1500	0.00913017121653591\\
1800	0.0107802799720774\\
2100	0.0123842152118321\\
2400	0.0139454724751415\\
2700	0.0154670264575544\\
3000	0.0169514451134956\\
};
\addplot[draw=black,mark=square,mark size=1.2] plot table[row sep=crcr] {%
300	0.00768362571013107\\
600	0.0144951443921183\\
900	0.0206088809399365\\
1200	0.0261517518226564\\
1500	0.0312190517735437\\
1800	0.0358841652413827\\
2100	0.0402048107206083\\
2400	0.0442272058354158\\
2700	0.0479889329090394\\
3000	0.0515209634251857\\
};
\legend{$\ell_1$,$\ell_2$,$\ell_3$}
\end{axis}
\end{tikzpicture}%
\hspace{-.65em}
%
\tikzsetnextfilename{aggprob_sysload_edge}
\begin{tikzpicture}

\begin{axis}[%
height=\figureheight,
width=\figurewidth,
at={(0.74in,0.469in)},
scale only axis,
separate axis lines,
every x tick label/.append style={font=\tiny, /pgf/number format/.cd, fixed, fixed zerofill, precision=1, /tikz/.cd},
every axis label/.style={font=\footnotesize},
xmajorgrids,
ymajorgrids,
xtick={300,1200,...,3000},
scaled x ticks=base 10:-3,
tick scale binop=\times,
every x tick scale label/.style={at={(axis cs:3000,-0.014)}},
xmin=0,
xmax=3300,
xlabel={system load},
every outer y axis line/.append style={black},
every y tick label/.append style={font=\tiny, /pgf/number format/.cd, fixed, fixed zerofill, precision=2, /tikz/.cd},
ytick={0,0.02,...,0.1},
ymin=0,
ymax=0.07,
scaled y ticks = false,
yticklabel pos=right,
axis background/.style={fill=white},
title={(b) edge caching},
reverse legend,
legend style={font=\tiny},
legend pos=north west,
]
\addplot[draw=black,mark=o,mark size=1.4] plot table[row sep=crcr] {%
300	0.000516722883906024\\
600	0.00101130938220822\\
900	0.00148899075792169\\
1200	0.00195299455780434\\
1500	0.00240548582999955\\
1800	0.00284800933146331\\
2100	0.00328172077609747\\
2400	0.00370751780751414\\
2700	0.00412611894874773\\
3000	0.0045381136519746\\
};
\addplot[draw=black,mark=x] plot table[row sep=crcr] {%
300	0.00230834512578331\\
600	0.00447283701469759\\
900	0.00652487412510341\\
1200	0.00848426897963666\\
1500	0.0103645835231455\\
1800	0.0121756745563516\\
2100	0.0139250426468983\\
2400	0.0156186052694415\\
2700	0.0172611679510465\\
3000	0.0188567259706889\\
};
\addplot[draw=black,mark=square,mark size=1.2] plot table[row sep=crcr] {%
300	0.0106972741240925\\
600	0.0194436143852784\\
900	0.0268394993201064\\
1200	0.0332442824845127\\
1500	0.0388905998067492\\
1800	0.0439378869242341\\
2100	0.04850025644644\\
2400	0.0526622223701918\\
2700	0.0564881331918193\\
3000	0.0600281165563905\\
};
\legend{$\ell_1$,$\ell_2$,$\ell_3$}
\end{axis}
\end{tikzpicture}%
  \caption{Impact of system load on the aggregation probability.}
  \label{fig:aggprob_sysload}
\end{figure}

\begin{figure}[h]
  \centering
  \setlength\figureheight{.155\textwidth} 
  \setlength\figurewidth{.192\textwidth}
%
%
%
%
\tikzsetnextfilename{aggperc_sysload_uniform}
\begin{tikzpicture}

\begin{axis}[%
height=\figureheight,
width=\figurewidth,
at={(0.74in,0.469in)},
scale only axis,
bar width=220,
separate axis lines,
every x tick label/.append style={font=\tiny, /pgf/number format/.cd, fixed, fixed zerofill, precision=1, /tikz/.cd},
every axis label/.style={font=\footnotesize},
ymajorgrids,
xtick={300,1200,...,3000},
scaled x ticks=base 10:-3,
tick scale binop=\times,
every x tick scale label/.style={at={(axis cs:3000,-1.6)}},
xmin=0,
xmax=3300,
xlabel={system load},
every outer y axis line/.append style={black},
every y tick label/.append style={font=\tiny},
ytick={0,2,...,14},
yticklabel=\pgfmathprintnumber{\tick}\%,
ymin=0,
ymax=8,
ylabel={aggregation percentage},
axis background/.style={fill=white},
title={(a) uniform caching},
reverse legend,
legend style={font=\tiny},
legend pos=north west,
]
\addplot[ybar stacked,draw=black,fill=black,area legend] 
plot table[y expr=\thisrow{y}, row sep=crcr] {%
x   y\\
300	0.0468666119576195\\
600	0.0926376654977009\\
900	0.137469003069999\\
1200	0.181476571716393\\
1500	0.224749633161554\\
1800	0.267358857934966\\
2100	0.309361518695688\\
2400	0.350804952834209\\
2700	0.39172894753876\\
3000	0.432167429265504\\
};
\addplot [color=black,solid,forget plot]
  table[row sep=crcr]{%
0	0\\
25	0\\
};
\addplot[ybar stacked,draw=black,fill=gray,area legend] 
plot table[y expr=\thisrow{y}, row sep=crcr] {%
x   y\\
300	0.173290784079301\\
600	0.33942105715759\\
900	0.499348150910285\\
1200	0.653792518077328\\
1500	0.803315440429447\\
1800	0.948366793863782\\
2100	1.0893157780863\\
2400	1.22647145363841\\
2700	1.36009691850211\\
3000	1.4904193712916\\
};
\addplot[ybar stacked,draw=black,fill=white,area legend] 
plot table[y expr=\thisrow{y}, row sep=crcr] {%
x   y\\
300	0.660187206280512\\
600	1.24351923923361\\
900	1.76534424343328\\
1200	2.23684099505175\\
1500	2.66640646960447\\
1800	3.06050958781362\\
2100	3.42424027149694\\
2400	3.76167575092078\\
2700	4.07613263922216\\
3000	4.37034505851344\\
};
\legend{$\ell_1$,$\ell_2$,$\ell_3$}
\end{axis}
\end{tikzpicture}%
\hspace{-.65em}
%
\tikzsetnextfilename{aggperc_sysload_edge}
\begin{tikzpicture}

\begin{axis}[%
height=\figureheight,
width=\figurewidth,
at={(0.74in,0.469in)},
scale only axis,
bar width=220,
separate axis lines,
every x tick label/.append style={font=\tiny, /pgf/number format/.cd, fixed, fixed zerofill, precision=1, /tikz/.cd},
every axis label/.style={font=\footnotesize},
ymajorgrids,
xtick={300,1200,...,3000},
scaled x ticks=base 10:-3,
tick scale binop=\times,
every x tick scale label/.style={at={(axis cs:3000,-1.6)}},
xmin=0,
xmax=3300,
xlabel={system load},
every outer y axis line/.append style={black},
every y tick label/.append style={font=\tiny},
ytick={0,2,...,14},
yticklabel=\pgfmathprintnumber{\tick}\%,
ymin=0,
ymax=8,
yticklabel pos=right,
axis background/.style={fill=white},
title={(b) edge caching},
reverse legend,
legend style={font=\tiny},
legend pos=north west,
]
\addplot[ybar stacked,draw=black,fill=black,area legend] 
plot table[y expr=\thisrow{y}, row sep=crcr] {%
x   y\\
300	0.0516722883906018\\
600	0.101130938220855\\
900	0.148899075792302\\
1200	0.195299455780427\\
1500	0.24054858299989\\
1800	0.284800933146411\\
2100	0.328172077609671\\
2400	0.370751780751331\\
2700	0.412611894874714\\
3000	0.45381136519735\\
};
\addplot [color=black,solid,forget plot]
  table[row sep=crcr]{%
0	0\\
25	0\\
};
\addplot[ybar stacked,draw=black,fill=gray,area legend]
plot table[y expr=\thisrow{y}, row sep=crcr] {%
x   y\\
300	0.202327272470754\\
600	0.39199323862364\\
900	0.571754021740663\\
1200	0.74334930461691\\
1500	0.907970321321601\\
1800	1.06648315478588\\
2100	1.21954708159751\\
2400	1.36768240875928\\
2700	1.5113118166478\\
3000	1.65078683550155\\
};
\addplot[ybar stacked,draw=black,fill=white,area legend] 
plot table[y expr=\thisrow{y}, row sep=crcr] {%
x   y\\
300	0.935455536878729\\
600	1.69638965923069\\
900	2.33651459497753\\
1200	2.8879860011122\\
1500	3.37162812437033\\
1800	3.80171758988525\\
2100	4.18847571541243\\
2400	4.53947460905358\\
2700	4.86048088401796\\
3000	5.15598749212163\\
};
\legend{$\ell_1$,$\ell_2$,$\ell_3$}
\end{axis}
\end{tikzpicture}%
  \caption{Impact of system load on cumulative aggregation
    percentage.}
  \label{fig:aggperc_sysload}
\vspace{-8pt}
\end{figure}
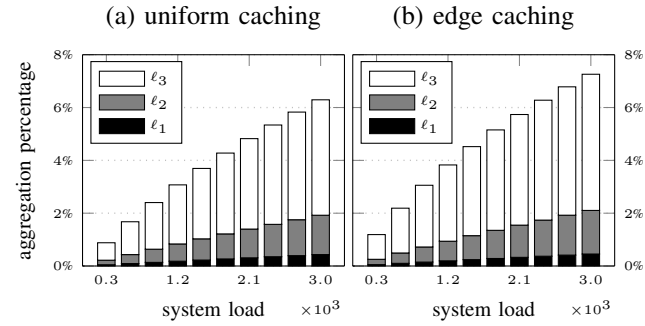

\fig{fig:aggprob_sysload} shows the probability of Interest
aggregation at each tree level as a function of system load. Contrary
to the results in \fig{fig:aggprob_delay}, a side-by-side comparison
of uniform- vs. edge-caching reveals that when the object catalog is
large, there is no remarkable difference between these two cache
allocation strategies. It is interesting that even with the highest
system load of 3000, the maximum aggregation probability observed at
the top most level is around 0.06. This almost 12-fold degradation
compared to the previous results highlights the importance of the size
of the object catalog in the overall odds of Interest aggregation.
This rather surprising finding can be explained as follows. With the
Zipf popularity distribution of objects, a highly popular object is
requested frequently. Therefore, once such an object is downloaded
into the CS, due to the frequent references to it, it stays there for
a long time. Hence, Interests for that object mostly result in cache
hits and are rarely aggregated. On the other hand, Interests for an
unpopular object (in the long tail of the distribution) are received
so sporadically over time that the odds of them co-occurring in the
short time span when the router is awaiting the content are almost
nil. As a result, in practice, Interest aggregation occurs only for a
small fraction of Interests.

To make this argument even stronger, \fig{fig:aggperc_sysload} shows
the cumulative percentage of aggregated Interests in the system
against an increasing system load. Evidently, the overall percentage
of Interests being aggregated is less than 5\% under a low to moderate
load, and around 7\% under heavy load. Note that for these results
each cache node has capacity to store only 0.07\% of the entire object
catalog.

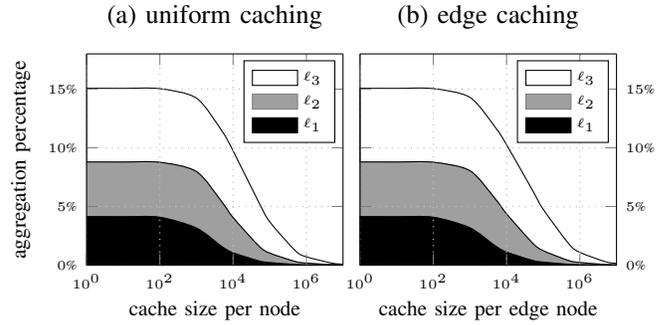
\begin{figure}[h]
  \centering
  \setlength\figureheight{.155\textwidth} 
  \setlength\figurewidth{.188\textwidth}
%
%
%
%
\tikzsetnextfilename{aggperc_cachesize_uniform}
\begin{tikzpicture}

\begin{axis}[%
width=\figurewidth,
height=\figureheight,
at={(0.731in,0.466in)},
scale only axis,
area style,
stack plots=y,
separate axis lines,
every x tick label/.append style={font=\tiny},
every axis label/.style={font=\footnotesize},
xmajorgrids,
ymajorgrids,
xmin=1,
xmax=1e7,
xtick={1,1e2,1e4,1e6},
xmode=log,
xlabel={cache size per node},
xlabel style={yshift=2pt},
every outer y axis line/.append style={black},
every y tick label/.append style={font=\tiny},
ymin=0,
ymax=18,
ytick={0,5,...,15},
yticklabel=\pgfmathprintnumber{\tick}\%,
ylabel={aggregation percentage},
axis background/.style={fill=white},
title={(a) uniform caching},
reverse legend,
legend style={font=\tiny},
legend pos=north east,
]
\addplot[fill=black,draw=black,smooth] plot table[row sep=crcr]{%
1	4.13681342767174\\
10	4.13246603304255\\
100	4.08548844017786\\
1000	3.16498689452457\\
5000	1.57807541464361\\
10000	1.06153099730909\\
50000	0.369411199804784\\
100000	0.224749633161554\\
500000	0.066052577552692\\
1000000	0.0379866560651651\\
5000000	0.00974185395064637\\
10000000	0.0051033518995325\\
}
\closedcycle;
\addplot[fill=darkgray,opacity=0.5,draw=black,smooth] plot table[row sep=crcr]{%
1	4.65543229808069\\
10	4.65719431084909\\
100	4.67728783437674\\
1000	4.81113795065896\\
5000	3.7901936872231\\
10000	2.99013839075576\\
50000	1.27400939618977\\
100000	0.803315440429447\\
500000	0.242931380426011\\
1000000	0.139617829826516\\
5000000	0.0347658336526331\\
10000000	0.0175367921776944\\
}
\closedcycle;
\addplot[fill=white,draw=black,smooth] plot table[row sep=crcr]{%
1	6.27502232514369\\
10	6.27478202991617\\
100	6.2724041245217\\
1000	6.25348251621477\\
5000	6.04420142313658\\
10000	5.68505245113703\\
50000	3.7262844302026\\
100000	2.66640646960447\\
500000	0.920303319532015\\
1000000	0.536417745270766\\
5000000	0.131044892610147\\
10000000	0.0637026371612518\\
}
\closedcycle;
\legend{$\ell_1$,$\ell_2$,$\ell_3$}
\end{axis}
\end{tikzpicture}
\hspace{-1.3em}
%
\tikzsetnextfilename{aggperc_cachesize_edge}
\begin{tikzpicture}

\begin{axis}[%
width=\figurewidth,
height=\figureheight,
at={(0.731in,0.466in)},
scale only axis,
area style,
stack plots=y,
separate axis lines,
every x tick label/.append style={font=\tiny},
every axis label/.style={font=\footnotesize},
xmajorgrids,
ymajorgrids,
xmin=1,
xmax=1e7,
xtick={1,1e2,1e4,1e6},
xmode=log,
xlabel={cache size per edge node},
xlabel style={yshift=2pt},
every outer y axis line/.append style={black},
every y tick label/.append style={font=\tiny},
ymin=0,
ymax=18,
ytick={0,5,...,15},
yticklabel=\pgfmathprintnumber{\tick}\%,
yticklabel pos=right,
axis background/.style={fill=white},
title={(b) edge caching},
reverse legend,
legend style={font=\tiny},
legend pos=north east,
]
\addplot[fill=black,draw=black,smooth] plot table[row sep=crcr]{%
1	4.13682729423998\\
10	4.13260509419229\\
100	4.08692499186336\\
1000	3.18466307447505\\
5000	1.62958326568259\\
10000	1.12095807718987\\
50000	0.415183005720996\\
100000	0.258819910778289\\
500000	0.0792772274818386\\
1000000	0.0461377083277464\\
5000000	0.0121662876103432\\
10000000	0.00651443608668349\\
}
\closedcycle;
\addplot[fill=darkgray,opacity=0.5,draw=black,smooth] plot table[row sep=crcr]{%
1	4.65547190971237\\
10	4.65759087374852\\
100	4.68130160318294\\
1000	4.86198090870916\\
5000	3.98506522103615\\
10000	3.24542191370466\\
50000	1.4976524708734\\
100000	0.972589137631187\\
500000	0.309431066793083\\
1000000	0.18063921963645\\
5000000	0.0469878167767821\\
10000000	0.0246434088391354\\
}
\closedcycle;
\addplot[fill=white,draw=black,smooth] plot table[row sep=crcr]{%
1	6.27502285545938\\
10	6.27478675943942\\
100	6.27238674707279\\
1000	6.24614497498763\\
5000	6.10107187459413\\
10000	5.89494974525652\\
50000	4.51294882741548\\
100000	3.52720635219959\\
500000	1.4256164133072\\
1000000	0.866481194995514\\
5000000	0.233117704917599\\
10000000	0.122761812821367\\
}
\closedcycle;
\legend{$\ell_1$,$\ell_2$,$\ell_3$}
\end{axis}
\end{tikzpicture}%
  \caption{Impact of cache size on overall aggregation percentage.}
  \label{fig:aggperc_cachesize}
\end{figure}

Increasing cache size further shrinks the benefit margin by improving
the overall cache hit rates. As \fig{fig:aggperc_cachesize} suggests,
with a small cache capacity, sizable gain (around 15\% total) can be
attained through Interest aggregation. However, this gain drastically
decays as cache size per node increases. Eventually, with a cache size
of 500,000 per node (\ie $<$ 0.4\% of the size of the content base),
there is virtually no benefit in Interest aggregation. A secondary
observation from the results in \fig{fig:aggperc_cachesize} is that
with smaller cache size, all layers in the hierarchy contribute about
the same in aggregation percentage; however, as more cache is added to
the nodes, most Interest aggregations occur at the upper layers, while
aggregation percentage at the edge approaches zero more rapidly.

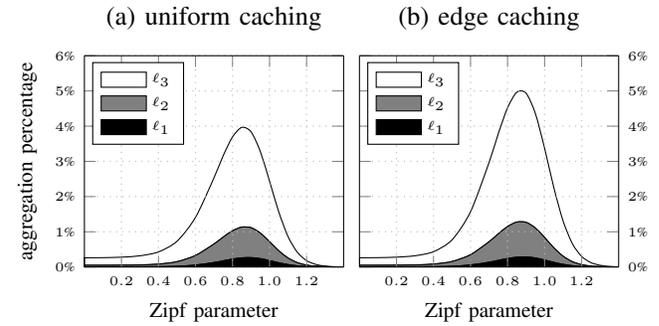
\begin{figure}[h]
  \centering
  \setlength\figureheight{.155\textwidth} 
  \setlength\figurewidth{.19\textwidth}
%
%
%
%
\tikzsetnextfilename{aggperc_alpha_uniform}
\begin{tikzpicture}

\begin{axis}[%
width=\figurewidth,
height=\figureheight,
at={(0.731in,0.466in)},
scale only axis,
area style,
stack plots=y,
separate axis lines,
every x tick label/.append style={font=\tiny, /pgf/number format/.cd, fixed, fixed zerofill, precision=1, /tikz/.cd},
every axis label/.style={font=\footnotesize},
xmajorgrids,
ymajorgrids,
xmin=0,
xmax=1.4,
xtick={0.2,0.4,...,1.2},
xlabel={Zipf parameter},
every outer y axis line/.append style={black},
every y tick label/.append style={font=\tiny},
ymin=0,
ymax=6,
ytick={0,1,...,6},
yticklabel=\pgfmathprintnumber{\tick}\%,
ylabel={aggregation percentage},
axis background/.style={fill=white},
title={(a) uniform caching},
reverse legend,
legend style={font=\tiny},
legend pos=north west,
]
\addplot[fill=black,draw=black,smooth] 
plot table[y expr=\thisrow{y}, row sep=crcr]{%
x   y\\
0	0.00641578413004131\\
0.1	0.00649561244292945\\
0.2	0.00684074953076888\\
0.3	0.00783558931155227\\
0.4	0.0109483838529492\\
0.5	0.0219046455947526\\
0.6	0.0551146756680107\\
0.7	0.127409617861167\\
0.75	0.176410206613011\\
0.8	0.224749633161554\\
0.85	0.259292957644316\\
0.9	0.266022098187905\\
0.95	0.237691169015734\\
1	0.182772530531448\\
1.05	0.120216458331498\\
1.1	0.0682752277986737\\
1.15	0.0340495895583351\\
1.2	0.0152871987872398\\
1.25	0.00632675812592354\\
1.3	0.00245826832416544\\
1.35	0.000912665415247904\\
1.4	0.000326752750168279\\
1.45	0.000114325742336053\\
1.5	3.90462949706308e-05\\
1.55	1.32188318629244e-05\\
1.6	4.38405620368613e-06\\
1.65	1.45059939464254e-06\\
1.7	4.72770624593303e-07\\
1.75	1.54445793345137e-07\\
1.8	4.98572271424904e-08\\
1.85	1.60188624899452e-08\\
1.9	5.11947920266569e-09\\
1.95	1.63402939499719e-09\\
2	5.21746704182982e-10\\
}
\closedcycle;
\addplot[fill=gray,draw=black,smooth] 
plot table[y expr=\thisrow{y}, row sep=crcr]{%
x   y\\
0	0.0427035441282114\\
0.1	0.0432329145643798\\
0.2	0.0455177352833127\\
0.3	0.0520475732444182\\
0.4	0.0714006920015198\\
0.5	0.129270916232666\\
0.6	0.272014892033543\\
0.7	0.52633134225084\\
0.75	0.674886587701224\\
0.8	0.803315440429447\\
0.85	0.87306645735394\\
0.9	0.85011847258832\\
0.95	0.725813318256852\\
1	0.536067630335172\\
1.05	0.339991802362608\\
1.1	0.186617375763665\\
1.15	0.0900656007903171\\
1.2	0.0391606842528764\\
1.25	0.0157002351736227\\
1.3	0.00591307478876861\\
1.35	0.00212883215542219\\
1.4	0.000739697263709646\\
1.45	0.000251232549922779\\
1.5	8.33928703764338e-05\\
1.55	2.74290316937113e-05\\
1.6	8.85348466269138e-06\\
1.65	2.84939803483519e-06\\
1.7	9.04642928414636e-07\\
1.75	2.87612050259565e-07\\
1.8	9.04768755577592e-08\\
1.85	2.83372322200297e-08\\
1.9	8.83238129017977e-09\\
1.95	2.74905904551924e-09\\
2	8.55630090776826e-10\\
}
\closedcycle;
\addplot[fill=white,draw=black,smooth] 
plot table[y expr=\thisrow{y}, row sep=crcr]{%
x   y\\
0	0.212856415943159\\
0.1	0.215476127449027\\
0.2	0.226746099616475\\
0.3	0.258479202321801\\
0.4	0.346718476683669\\
0.5	0.579528258923369\\
0.6	1.08199581154329\\
0.7	1.88128565339306\\
0.75	2.31410113205288\\
0.8	2.66640646960447\\
0.85	2.83239491947073\\
0.9	2.72260063484433\\
0.95	2.31851459259368\\
1	1.72067057358605\\
1.05	1.10106662023785\\
1.1	0.609178301197335\\
1.15	0.295131735276217\\
1.2	0.128014814004367\\
1.25	0.0508880914508251\\
1.3	0.0189163682046172\\
1.35	0.00669632996607018\\
1.4	0.00228322556956203\\
1.45	0.000759109555306635\\
1.5	0.000246659159477851\\
1.55	7.9223100670728e-05\\
1.6	2.50124027681367e-05\\
1.65	7.85611775630412e-06\\
1.7	2.4380396509845e-06\\
1.75	7.55793979595982e-07\\
1.8	2.32179807223677e-07\\
1.85	7.100626269652e-08\\
1.9	2.16163753534971e-08\\
1.95	6.56734960214572e-09\\
2	1.99342836138948e-09\\
}
\closedcycle;
\legend{$\ell_1$,$\ell_2$,$\ell_3$}
\end{axis}
\end{tikzpicture}
\hspace{-.5em}
%
\tikzsetnextfilename{aggperc_alpha_edge}
\begin{tikzpicture}

\begin{axis}[%
width=\figurewidth,
height=\figureheight,
at={(0.731in,0.466in)},
scale only axis,
area style,
stack plots=y,
separate axis lines,
every x tick label/.append style={font=\tiny, /pgf/number format/.cd, fixed, fixed zerofill, precision=1, /tikz/.cd},
every axis label/.style={font=\footnotesize},
xmajorgrids,
ymajorgrids,
xmin=0,
xmax=1.4,
xtick={0.2,0.4,...,1.2},
xlabel={Zipf parameter},
every outer y axis line/.append style={black},
every y tick label/.append style={font=\tiny},
ymin=0,
ymax=6,
ytick={0,1,...,6},
yticklabel=\pgfmathprintnumber{\tick}\%,
yticklabel pos=right,
axis background/.style={fill=white},
title={(b) edge caching},
reverse legend,
legend style={font=\tiny},
legend pos=north west,
]
\addplot[fill=black,draw=black,smooth] 
plot table[y expr=\thisrow{y}, row sep=crcr]{%
x   y\\
0	0.00641986025299073\\
0.1	0.0064998564358634\\
0.2	0.0068459502788135\\
0.3	0.00784673360771292\\
0.4	0.0110296252084399\\
0.5	0.022545362206668\\
0.6	0.0580423842594792\\
0.7	0.135696361828411\\
0.75	0.188405469369658\\
0.8	0.24054858299989\\
0.85	0.278048724571334\\
0.9	0.285788480019714\\
0.95	0.256771900471087\\
1	0.198933879338978\\
1.05	0.131484637089099\\
1.1	0.0752551972398079\\
1.15	0.0377693220731759\\
1.2	0.0170469778302038\\
1.25	0.00708772163955042\\
1.3	0.00276990310440514\\
1.35	0.00103184596806519\\
1.4	0.000369982803312788\\
1.45	0.000129555109310868\\
1.5	4.45511923271289e-05\\
1.55	1.49999364197851e-05\\
1.6	5.00115751383757e-06\\
1.65	1.65496275334781e-06\\
1.7	5.42888254646308e-07\\
1.75	1.76247568918569e-07\\
1.8	5.69742389986732e-08\\
1.85	1.83791158791084e-08\\
1.9	5.88417730680113e-09\\
1.95	1.88385109459981e-09\\
2	5.99570771213427e-10\\
}
\closedcycle;
\addplot[fill=gray,draw=black,smooth] 
plot table[y expr=\thisrow{y}, row sep=crcr]{%
x   y\\
0	0.0427425825299923\\
0.1	0.0432735585376303\\
0.2	0.045567512462277\\
0.3	0.0521527871832157\\
0.4	0.0720709325788423\\
0.5	0.133898082594097\\
0.6	0.291832943786146\\
0.7	0.581250596967654\\
0.75	0.754272291769919\\
0.8	0.907970321321601\\
0.85	0.997727565382552\\
0.9	0.982375234637248\\
0.95	0.851300500883848\\
1	0.639809950128258\\
1.05	0.412115622299025\\
1.1	0.230538593505513\\
1.15	0.113289435572421\\
1.2	0.0501139892495999\\
1.25	0.0204340846372299\\
1.3	0.00783640094436122\\
1.35	0.00286669115407146\\
1.4	0.00101021962054632\\
1.45	0.000347964905339497\\
1.5	0.000117808219020646\\
1.55	3.9086275790622e-05\\
1.6	1.2852568533134e-05\\
1.65	4.197950745524e-06\\
1.7	1.36022751667979e-06\\
1.75	4.36490985288589e-07\\
1.8	1.39559157677251e-07\\
1.85	4.4554115130831e-08\\
1.9	1.41243610235471e-08\\
1.95	4.47989918701359e-09\\
2	1.41319921311314e-09\\
}
\closedcycle;
\addplot[fill=white,draw=black,smooth] 
plot table[y expr=\thisrow{y}, row sep=crcr]{%
x   y\\
0	0.213279157623423\\
0.1	0.215916196527532\\
0.2	0.227284346184405\\
0.3	0.259587561074035\\
0.4	0.352670229553159\\
0.5	0.614794180914466\\
0.6	1.22092478236375\\
0.7	2.25155574668322\\
0.75	2.84636050725067\\
0.8	3.37162812437033\\
0.85	3.68912968550102\\
0.9	3.6664835582082\\
0.95	3.25264595407706\\
1	2.53414367856518\\
1.05	1.7075541149656\\
1.1	1.00024791230719\\
1.15	0.512428593788016\\
1.2	0.234327735833361\\
1.25	0.0978774767458647\\
1.3	0.0381525752260892\\
1.35	0.0141043723865453\\
1.4	0.00500303024708583\\
1.45	0.00173007729538854\\
1.5	0.000587113853162069\\
1.55	0.000195062045800177\\
1.6	6.41929183687026e-05\\
1.65	2.09766324244258e-05\\
1.7	6.79870363238185e-06\\
1.75	2.18201004573038e-06\\
1.8	6.97714853540555e-07\\
1.85	2.22755871546279e-07\\
1.9	7.06191626045462e-08\\
1.95	2.23990202366521e-08\\
2	7.06591115553498e-09\\
}
\closedcycle;
\legend{$\ell_1$,$\ell_2$,$\ell_3$}
\end{axis}

\end{tikzpicture}%
  \caption{Impact of popularity distribution on aggregation
    percentage.}
  \label{fig:aggperc_alpha}
\vspace{-8pt}
\end{figure}

Finally, \fig{fig:aggperc_alpha} captures the impact of the object
popularity distribution on the cumulative percentage of aggregated
Interests. The non-monotonic trend of curves in
\fig{fig:aggperc_alpha} exhibits a diminishing returns type of
effect. To explain this behavior, we note that with a Zipf popularity
distribution, objects can heuristically be categorized into two
groups, namely, an unpopular majority and a popular minority. The Zipf
parameter ($\alpha$) controls the relative size of each group as well
as the skewness of the distribution. In fact, the larger the Zipf
parameter, the smaller the proportion of the minority group, and the
greater their popularity intensity. The latter signifies a higher
access frequency to the objects in the popular group as $\alpha$
increases, hence a higher aggregation rate for them. However, as
$\alpha$ increases and the proportion of the popular objects shrinks,
the higher access frequency also results in higher hit rates, because
a lot of those objects eventually find their way into the caches;
therefore, subsequent requests for them no longer get aggregated. On
the other hand, thanks to their diminishing popularity, the Interests
for the majority group are also becoming so sparsely dispersed that
the odds of finding a relevant entry for them at the PIT becomes
negligible. For this, only a small fraction of Interests representing
those fairly popular objects which may not have found a free spot in
(limited-size) caches remain subject to aggregation. Further
increasing the Zipf parameter shrinks down the size of the popular
group gradually such that at some point, every one of them finds a
permanent place in all caches. Thenceforth, the probability of
aggregation becomes effectively zero.

From another viewpoint, \fig{fig:aggperc_alpha} also provides
suggestive evidence that even under a non-stationary content
popularity distribution, no remarkable benefit can be anticipated from
Interest aggregation. For example, when object references are
temporally localized, a data object becomes highly popular over a
certain duration of time, while its popularity gradually vanishes over
time as some other data object becomes popular. In that case, if the
objects popularity is measured within smaller discrete time windows,
each piece can independently be approximated with a Zipf distribution
with a possibly different parameter. As \fig{fig:aggperc_alpha}
suggests, irrespective of how different the popularity profile looks
like, only an insignificant number of Interests may be aggregated at
various intervals; hence, the benefits of Interest aggregation would
still remain minimal, with the maximum aggregation of less than 5\%
taking place around a Zipf parameter of 0.9.


\section{Final Remarks and Conclusions} \label{sec:conclusions}

We presented the first analytical treatment of Interest aggregation in
Content-Centric Networks using a simple yet accurate model where
content download delays into the routers are non-zero. Based on our
model, we introduced an iterative algorithm for analyzing a
hierarchical network of content routers in terms of CS hit- and PIT
aggregation probabilities and router response times. This method
enables the evaluation of large-scale hierarchical caching structures,
such as that of an ICN at Internet scale, with high accuracy and low
computational cost for which discrete-event simulations are entirely
impractical due to high processing and time demands.

Our numerical evaluations of a network of caches under realistic
assumptions revealed that: (1) even with very small caching budgets,
less than 5\% of total Interests on average are subject to
aggregation; (2) increasing caching budgets rapidly diminishes the
benefits of Interest aggregation; (3) most aggregations take place
closer to the producers, negating the expected benefits of reducing
latency and bandwidth utilization desired from aggregation; and (4)
aggregation gains are almost invariant to the choice of cache
allocation strategy (\ie edge- vs. uniform-caching). Together, these
observations imply that Interest aggregation should only be an
optional mechanism in Content-Centric Networking. Furthermore, if
per-Interest forwarding state is not needed for other purposes, the
statefull forwarding plane of NDN (realized through PITs) can
effectively be replaced with more efficient mechanisms, such as
CCN-DART~\cite{jj-icccn:15,jj-icnc:16} and CCN-GRAM
\cite{jj-ifip:16}, in which forwarding state is stored only per route
or per destination while providing similar end-to-end content delivery
latencies.

Our model relies on the assumption that 
input streams conform to the independent reference model, which need
not be true in reality.  
However, the simulation results in \cite{jj-icnc:16, jj-ifip:16}
indicate that in-network caching makes Interest aggregation
unnecessary even with spatio-temporal locality of Interests.

\section*{Acknowledgments}

This work was supported in part by the Jack Baskin Chair of Computer
Engineering at UCSC, NSF grant CNS-1413998, and MURI ARO grant
W911NF-12-10385.


\bibliographystyle{sty/IEEEtran} 
\balance
\bibliography{bib/biblio.tex} 

\end{document}